\documentclass{mem}
\usepackage{natbib}\usepackage{txfonts}\usepackage{balance}
\usepackage{graphicx}
\usepackage[a4paper,breaklinks,dvipdfm]{hyperref}
\idline{75}{282}
\begin{document}
\def\teff{$T\rm_{eff }$}
\def\kms{$\mathrm {km s}^{-1}$}
\def\Ms{M$_\odot$}

\title{Star clusters as tracers of galactic nuclei properties
}

\subtitle{}

\author{M. \,Arca-Sedda\inst{1} and R. \, Capuzzo-Dolcetta\inst{1} }

\institute{Universit\'a di Roma ``Sapienza'', Dipartimento di Fisica,
P.le Aldo Moro, 5,
I-00185 Roma
\email{m.arcasedda@gmail.com}
}

\authorrunning{M. Arca-Sedda and R. Capuzzo-Dolcetta}
\titlerunning{Star clusters and galactic nuclei}

\abstract{
We present a series of $N$-body simulations representing the evolution of a galactic nucleus and its stellar content in a nearly one-to-one representation. The aim of this suite of simulations is to shed light on the interplay between nuclear clusters (NCs), super-massive black holes (SMBH) and the galactic nuclei in which they are contained.
We modelled galaxies with masses from few times $10^8$ to $10^{11}$ \Ms, hosting in their nucleus a number of globular clusters and, in some cases, a central SMBH.
}

\maketitle{}

\section{Introduction}
The innermost region of many galaxies is dominated by the presence of a massive and dense cluster of stars, usually called nuclear cluster (NC), quite independently on the galaxy type or age \citep{cote06}. These compact systems oftenly contain at their centre a super-massive black hole (SMBH). Our current understanding of galactic nuclei suggests that the mass of the galaxy determines the kind of compact massive object (CMO) contained within its nucleus. Indeed, galactic nuclei heavier than $10^{11}$ \Ms are dominated by SMBHs, while NCs seem to be preferentially located in galaxies with masses $\lesssim 10^{9}$ \Ms, although they are poorly observed in dwarf galaxies ($M_g\simeq 10^8$ \Ms). The two objects co-exist in galaxies with masses in the range $10^9-10^{11}$ \Ms.
The existence of observational correlations between NCs and their host galaxies helped in constraining the possible formation scenarios for these object, and relate them to SMBHs formation and evolution \citep{LGH,Ant13,ASCD14df}. According to the ``{\it dry-merger}'' formation scenario, 
NCs would originate by the collision and merging of orbitally segregated globular clusters (GCs), which inspiral toward the galactic nucleus as a consequence of dynamical friction (df) \citep{Trem76,Dolc93}. 

During the infall process, the GCs decay is opposed by the tidal forces exerted by the galactic nucleus and the SMBH. This process is called tidal heating (th). 
Studying the competing action of th and df through numerical simulations, \cite{ASCDS16} have shown that NCs and SMBHs likely follow different formation pathways, but their subsequent co-evolution is strictly connected. 

In particular, we can divide galaxies in df dominated, where df acts efficiently on the GCs motion and NCs form easily, and th dominated, where tidal forces exerted from the galactic background and the SMBH disrupt the infalling GCs, preventing NC formation \citep{ASCDS16}. 

Direct $N$-body simulations are perfect tools for studying the GC orbital decay and merging and, possibly, testing the formation and evolution of NCs \citep{DoMioA,AMB,ASCD15he}. 
However, the extremely high computational cost of these simulations limited so far the number of bodies to few thousands, making infeasible the modelling of a galactic nucleus. 
The recent achievements of high performance computing allow us today to run $N$-body simulations on hybrid computers comprised of ordinary central processing units (CPUs) and graphic processing units (GPUs), which can speed up the calculations by a factor 100. This implies the possibility to simulate the inner region of a galaxy with an unprecedented level of detail \cite{ASCD15he}. 
Moreover, the accuracy of modern observations give us the possibility to get detailed initial conditions for numerical models, thus maximising their reliability.

In this contribution, we present a suite of simulations of galactic nuclei carried out in the framework of the project ``MEGaN: modelling the evolution of galactic nuclei''. Varying galaxy masses and properties, we test the dry-merging scenario in both df and th dominated systems, simulating the evolution of a galactic nucleus in a nearly one-to-one representation. We will focus on three different systems: a dwarf galaxy model, based on data related to the Fornax dwarf spheroidal galaxy (dSph), with mass $1.5 \times 10^8$ \Ms and 5 old GCs with masses below $5\times 10^5$ \Ms; a medium sized galaxy model, based on available data for the Henize 2-10 dwarf starburst galaxy ($M\simeq 1.6\times 10^9$ \Ms), containing a $2\times 10^6$ \Ms SMBH and 11 young massive clusters (YMCs); a large galaxy model, with a mass $10^{11}$ \Ms, containing a SMBH with mass $10^8$ \Ms and 42 GCs with masses up to $2\times 10^6$ \Ms.

\section{The numerical approach}

We discuss here a series of numerical modelling of both df and th dominated galaxies. Our models consist of three main ingredients: a host galaxy, a central MBH, which may be present or not, and a sample of star clusters. 

To represent an entire galaxy, our models should consist of at least $10^8-10^9$ particles, and up to $10^{11}$, a number exceedingly large even for state-of-art computational resources. However, our aims are to investigate the evolution of the galactic nucleus only, then we can limit our models to this region. Due to this, in all the cases considered, we modelled the host galaxy according to a truncated Dehnen density profile \citep{deh, ASCD15he}:
\begin{equation}
\rho(r) = \rho_g\left(\frac{r}{r_g}\right)^{-\gamma}\left(\frac{r}{r_g}+1\right)^{-4+\gamma}{\mathrm cosh}\left(\frac{r}{r_{\rm cut}}\right),
\end{equation}
where $\gamma$ is the slope of the density profile, $\rho_g=(3-\gamma)M_g/(4\pi r_g^3)$ is the galaxy typical density, $M_g$ its total mass and $r_g$ its scale radius.  
The truncation radius, $r_{\rm cut}$, allow us to build self-consistent systems over length-scale consistent with the galaxy nuclei. This allows to model only a limited region of the galaxy, thus reducing the number of particles needed to represent the galaxy.  

The star clusters properties, instead, are sampled according to King profiles \citep{King}. In this case, the choice of parameters is constrained by observations, as we will discuss in the following.

We carried out our simulations using the \texttt{HiGPUs} direct $N$-body code \citep{Spera}, which run on GPU servers and allowed us to model the whole galactic nucleus and its star clusters with more than $10^6$ particles. This implies that each particle in our simulations has a mass $\sim 10-80$ M$_\odot$, depending on the simulation considered. \texttt{HiGPUs} does not contain any recipe for treating hard encounters, thus forcing us to soften the gravitational interactions through a smooting length $\epsilon \lesssim 0.05$ pc.

\section{Results}

In this section, we briefly summarize some of the results achieved through the suite of simulations performed. In particular, we will discuss the possible formation of NCs in dwarf, mid-sized and heavy galaxies.
\begin{figure*}[t!]
\resizebox{\hsize}{!}{
\includegraphics[clip=true]{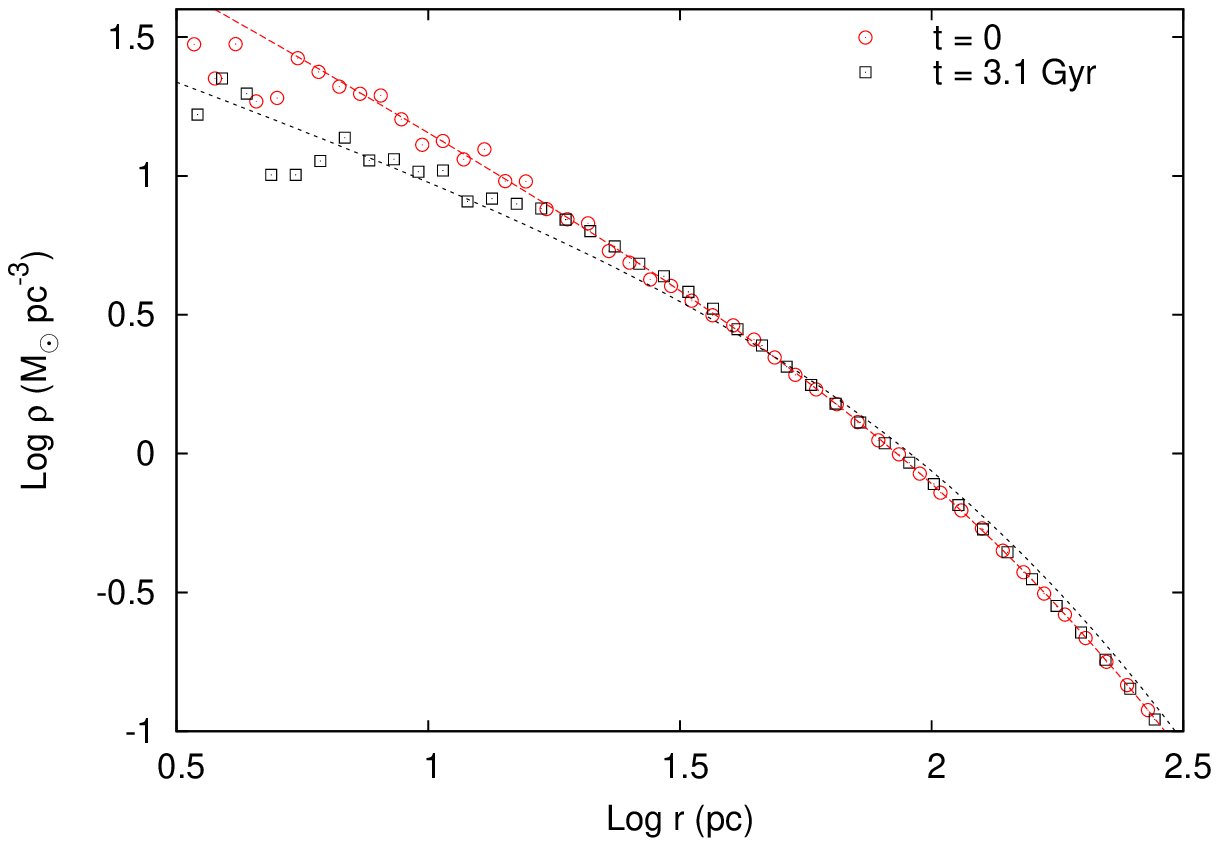}
\includegraphics[clip=true]{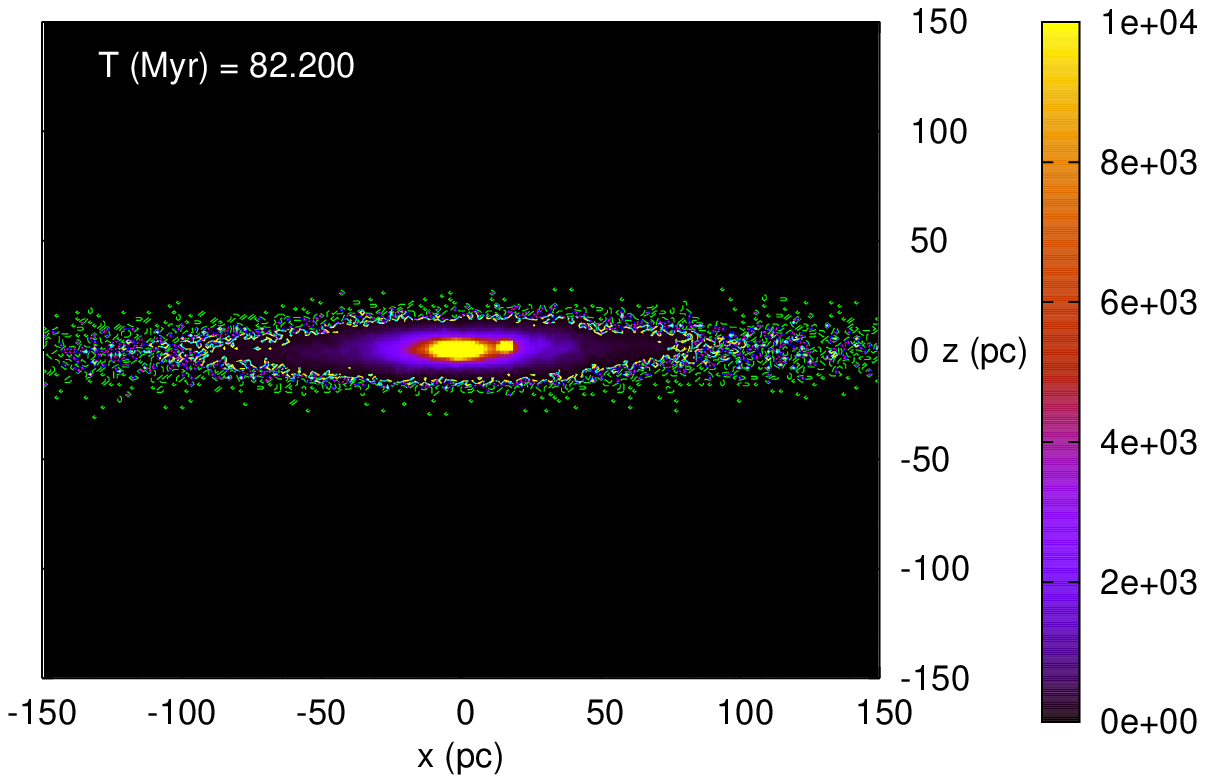}
\includegraphics[clip=true]{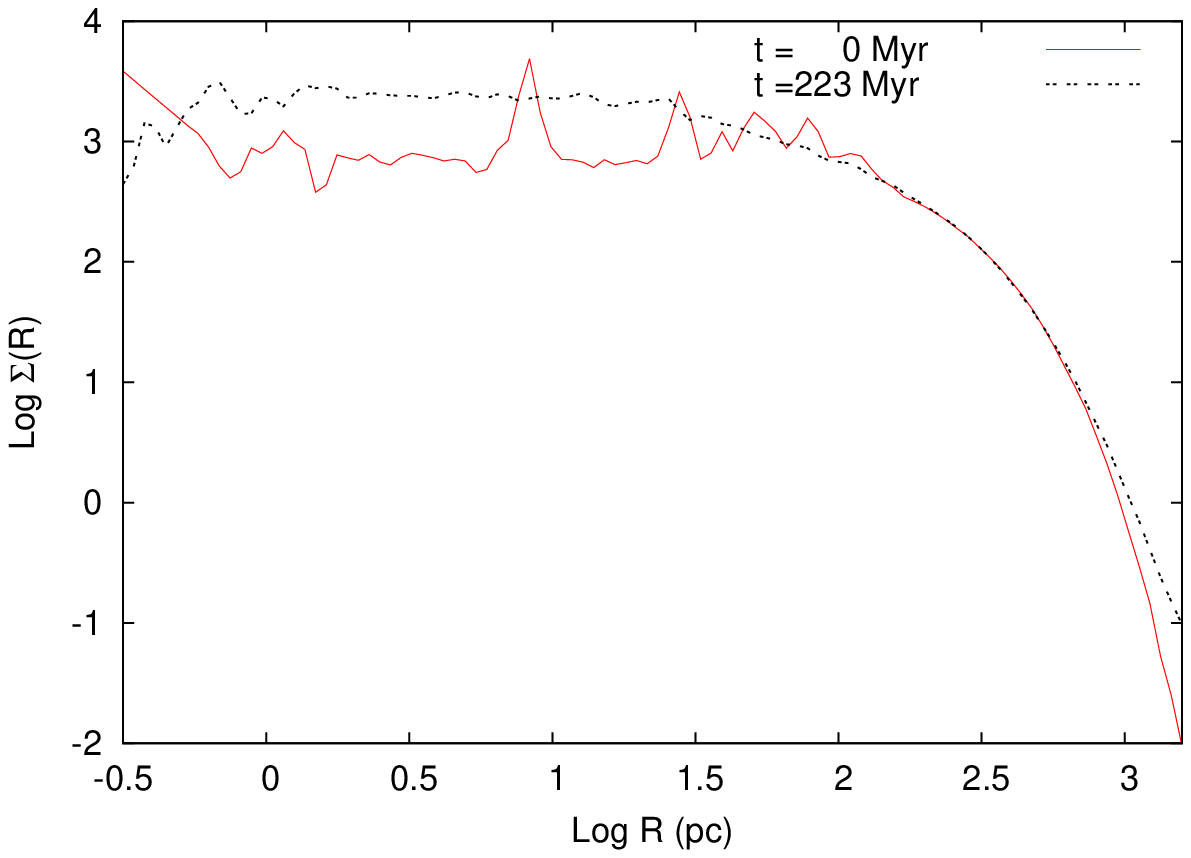}
}
\caption{\footnotesize
Left panel: evolution of the Fornax density profile in one of the simulations carried out. It is evident that it flattens after $3.2$ Gyr. The final slope, after the orbital decay and disruption of the Fornax GCs, is $\sim 0.6$. Central panel: surface map (on the xz plane) of one of the Henize 2-10 simulations performed. The GCs initial positions lie all in the xy plane. Their infall drives the formation of a disc with properties quite similar to those of observed NDs. Right panel: surface density profile of the $10^{11}$ \Ms galaxy model. The absence of a bright central overdensity implies that the NC formation is completely suppressed by tidal forces ,as expected for a th dominated system.}
\label{F1}
\end{figure*}

\subsection{Dwarf galaxies: GCs inspiral and the cusp/core problem}

Our Milky Way (MW) hosts in its surroundings a number of dSph galaxies, with masses few times $10^8$ M$\odot$. It is widely believed that dSph are contained within massive dark matter halos (DMH). However, several authors have pointed out that typical dSphs' density profiles are significantly flatter than what expected from the $\Lambda$-CDM paradigm. The Fornax dSph is the most massive dwarf among the MW satellites, and hosts 5 old globular clusters. We run several $N$-body simulations of the Fornax journey around the MW, and the orbital evolution of the 5 GCs. We assumed two different density profiles for Fornax: i) $\rho(r)\propto r^{-1}$, as suggested by standard CDM model, and ii) $\rho(r)\propto r^{-1/2}$, as suggested by observations. We found that in the flatter case, GCs decay lead to the formation of an evident bright nucleus, very similar to the observed NCs. This is the first simulation in which the dry-merger scenario is tested in a galaxy that does not contain any SMBH in its centre.
On the other hand, when a steeper density profile is considered, we found that th dominates, disrupting the infalling GCs and preventing NC formation. We found that the GCs remnants exert a strong feedback on the background, leading to a significant flattening of the density profile, as shown in Fig. \ref{F1}. It is worth noting that the slope of the $\rho(r)$ function passes from $-1$ to $-0.6$, a value quite close to what expected from observations. Our results suggest that the orbital decay of star clusters in the early life of a dSph can efficiently erase the initial DM cusp. On the other hand, the possibility that dSphs have nearly flat density profiles at their birth seem to be hardly compatible with the observational dearth of NCs in these small systems.

\subsection{Middle size galaxies: the Henize 2-10 dwarf starburst}

The starburst galaxy Henize 2-10 is a dwarf starburst galaxy located in the Pisces constellation. Its nucleus host a central MBH with mass $2.6\times 10^6$ \Ms \citep{reines12} and 11 super-star clusters (SSCs) \citep{ngu14}, with ages below $10$ Myr. The contemporary presence of SSCs and a MBH makes Henize 2-10 an ideal laboratory for testing the dry-merger scenario on a strong observational basis. Moreover, the similarity between the Henize 2-10 and the MW MBHs give us the opportunity to put further constraints on the formation of our own NC. Varying the SSCs orbital parameters, we have shown that their infall leads to the formation of a NC in any case \citep{ASCD15he}. We found that the galactic background affects significantly the NC properties and final mass, as well as the MBH does. In particular, while the former regulates the disruption of low-mass clusters, the latter determines the amount of mass brought to the forming NC from the most massive clusters, that penetrate deeper into the galactic nucleus. Moreover, we have shown that the decay and merging of the SSCs can lead to the formation of a disky structure (see Fig. \ref{F1}). This structure has structural properties quite similar to nuclear stellar discs (NDs), massive discs of stars observed in many galaxies of every Hubble type in place of NCs. Hence, our results suggest that the dry-merger scenario can represent a valid formation channel for both NCs and NDs.

\subsection{Heavy galaxies: quenching NCs formation}

The absence of NCs in heavy galaxies has been ascribed to the intense tidal forces exerted from the SMBH that reside in their centre \citep{Ant13,ASCD14df,ASCDS16}. However, none of these works tested this possibility modelling the galactic nucleus and its globular cluster system (GCS). To investigate the validity of previous findings, we modelled the evolution of 42 massive GCs masses between $3\times 10^5$ and $2\times 10^6$ \Ms, traversing the nucleus of a galaxy with a total mass $M=10^{11}$ \Ms and hosting a SMBH with mass $10^8$ \Ms. As suggested by previous authors, such value for the SMBH mass should represent the treshold above which NCs formation is completely quenched. We found that after $\gtrsim 200$ Myr, the $80\%$ of them have been completely disrupted by the tidal forces. As a consequence, no evident nuclei can be observed in the projected density profile, as shown in Fig. \ref{F1}. The huge amount of data produced (over $3$ Tb) allowed us to investigate a number of phenomena occurring in these dense regions. For instance, we followed the evolution of the GCs mass function, $F(M)$. We found that an initially flat distribution of mass evolves toward a power-law function with slope $F(M)\propto-0.78$. Another interesting phenomena is the ejection of stars at high velocities, which occur as a consequence of close encounters between a GC and a SMBH. According to previous results, we found that $\sim 5\%$ of all the stars (GCs + host galaxy) are thrown away from the nucleus with velocities up to $3\times 10^2$ km s$^{-1}$. Rescaling the results to a real galaxy, and assuming for the stars a Kroupa IMF, which implies a mean stellar mass $0.62$ \Ms, we would expect that more than $10^6$ ejected stars would traverse its outer regions with velocities above $200$ km s$^{-1}$, reaching in few cases values above $10^3$ km s$^{-1}$. The fraction of stars ejected with lower velocities, instead, constitutes the $1\%$ of the total mass of the system, thus suggesting that this mechanism can represent an interesting process for enriching the population of stars moving in the galactic halo.


\begin{acknowledgements}
MAS acknowledges financial support from the University of
Rome ``Sapienza'' through the grant ``52/2015'' in the frame-
work of the research project ``MEGaN: modelling the envi-
ronment of galactic nuclei''
\end{acknowledgements}

\footnotesize{
\bibliographystyle{aa}
\bibliography{bblgrphy}
}

\end{document}